\documentclass[12pt,preprint]{aastex}

\shorttitle{X-ray Evolution of PWNe}
\shortauthors{Bamba et al.}

\begin{document}


\title{X-ray Evolution of Pulsar Wind Nebulae}

\author{
Aya Bamba\altaffilmark{1,2},
Takayasu Anada\altaffilmark{2},
Tadayasu Dotani\altaffilmark{2},
Koji Mori\altaffilmark{3},
Ryo Yamazaki\altaffilmark{4},
Ken Ebisawa\altaffilmark{2},
Jacco Vink\altaffilmark{5}
}

\altaffiltext{1}{
School of Cosmic Physics, Dublin Institute for Advanced Studies
31 Fitzwilliam Place, Dublin 2,
Ireland
abamba@cp.dias.ie
}

\altaffiltext{2}{
ISAS/JAXA Department of High Energy Astrophysics
3-1-1 Yoshinodai, Chuo-ku, Sagamihara,
Kanagawa 252-5210, JAPAN
}

\altaffiltext{3}{
Department of Applied Physics, Faculty of Engineering
University of Miyazaki,
1-1 Gakuen Kibana-dai Nishi, Miyazaki, 889-2192, Japan
}

\altaffiltext{4}{
Department of Physics and Mathmatics, 
Aoyama Gakuin University,
5-10-1 Fuchinobe, Sagamihara, Kanagawa, 252-5258, Japan
}

\altaffiltext{5}{
Astronomical Institute, Utrecht University,
P.O. Box 80000, 3508TA Utrecht, The Netherlands
}

\begin{abstract}
During the search for counterparts of very-high-energy
gamma-ray sources,
we serendipitously discovered large, extended, low surface brightness
emission from PWNe
around pulsars with the ages up to $\sim$100~kyrs,
a discovery made possible by the low and stable background of
the {\it Suzaku} X-ray satellite.
A systematic study of a sample of 8 of these PWNe,
together with {\it Chandra} datasets,
has revealed us that
the nebulae keep expanding up to for $\sim$100~kyrs,
although time scale of the synchrotron X-ray emission is only $\sim$60~yr
for typical magnetic fields of 100~$\mu$G.
Our result suggests that the accelerated electrons up to $\sim$80~TeV
can escape from the PWNe without losing most energies.
Moreover, in order to explain the observed correlation between
the X-ray size and the pulsar spindwon age,
the magnetic field strength in the PWNe must decrease with time.
\end{abstract}


\section{Introduction}

Pulsar Wind Nebulae (PWNe) 
consist of electrons and positrons
accelerated by the pulsar wind termination shocks,
giving rise to bright X-ray synchrotron emission
from young PWNe with ages up to
several kyr.
Recent X-ray observations have shown 
the presence of the extended PWNe
\citep[e.g.,][]{camilo2004,romani2005,vanetten2008}
from the pulsars with the age of up to $\sim$10 kyr.
However, we still do not know how the PWNe evolve
when they become older than the energy-loss time scales of electrons,
and whether the accelerated electrons can escape the PWNe or not.
Understanding the electron/positron escape process is important,
since pulsars and PWNe may be an %origin of
important source of 
cosmic ray electrons and positrons \citep{kawanaka2010},
as recently suggested by
PAMELA \citep{adriani2009}, ATIC \citep{chang2008},
PPB-BETS \citep{torii2008}, Fermi \citep{abdo2009},
and H.E.S.S. \citep{aharonian2009}.

Recently, the Galactic plane survey by H.E.S.S. Cherenkov telescopes
has revealed about 50 
very-high-energy (VHE) gamma-ray sources
\citep{aharonian2005,aharonian2006,aharonian2008}.
%Since there are old pulsars in the vicinities, 
%some of the gamma-ray sources 
%are now categorized as old PWNe although
%their extended emission have not been found in other wavelengths.
The search for the counterparts in other wavelengths
is ongoing.
In fact,
%Actually, 
about a half of such VHE gamma-ray sources without 
any counterparts are now categorized as PWNe\footnote{%
see http://tevcat.uchicago.edu/\ .},
because possibly associated old X-ray pulsars are found
later.
%{\bf
%~(Comment: Why the sources WITHOUT ANY COUNTERPARTS
%can be classified as PWNe? 
%So, I would suggest:
%Furthermore, for about a half of 
%galactic VHE gamma-ray sources, the possible origin is
%said to be the PWN.
%)~
%}
Hence these sources
are keys to understand the PWN evolution
and how the accelerated particles escape from the system.

However, the problem is that such old PWNe are rather faint in X-rays
as opposed to in VHE gamma-rays
\citep{mattana2009}.
As the PWNe ages, the X-ray emission becomes rapidly fainter,
while the VHE gamma-ray luminosity keeps constant.
Previous X-ray observations did not have
enough sensitivity to detect the probable faint nebulae.
We thus need deep X-ray observations with low background
to understand their X-ray properties.
The X-ray Imaging Spectrometer \citep[XIS;][]{koyama2007}
onboard {\it Suzaku} \citep{mitsuda2007} has carried out deep follow-ups
for these sources to search for possible X-ray counterparts.
Actually, \citet{uchiyama2009} found a largely extended nebula
from PSR~B1823-13 associated with HESS~J1825$-$137,
which could not be detected
with {\it Chandra} \citep{pavlov2008}
or {\it XMM-Newton} \citep{gaensler2003} observations.
PSR~J1809$-$1917 is also in the same case
with {\it Chandra} \citep{kargaltsev2007} and {\it Suzaku} \citep{anada2010}.
In this paper, we will make a systematic study of X-ray structure of 
such old PWNe.

\section{Sample Selection and Results}

The XIS has a low and stable background,
and is ideal for observing faint and extended sources like our targets.
We selected the XIS deep observations of the 3 PWNe
with VHE gamma-ray counterparts
(HESS~J1825$-$137, HESS~J1809-193, HESS~J1718$-$385)
and pulsars with known characteristic age ($t_c = P/(2\dot{P})$, with
$P$ and $\dot{P}$ period and period derivative, respectively).
For comparison, 
we also added a sample of five younger pulsars with VHE gamma-rays
that have been observed with {\it Chandra}.
Vela is not used in this paper although it is a well known 
PWN with TeV emission,
since it is too extended to cover by one or a few pointings of
current X-ray observatories.
Table~\ref{tab:result} lists the observational details
and physical parameters of our samples,
which cover most of the PWNe
with known periods and period derivatives, distances,
and archival deep X-ray observations.
Our sample set covers the PWN candidates with ages
in the range of $\sim$kyr up to $\sim$100~kyr. 

\begin{figure}
\epsscale{.3}
%\plotone{kes75_v1.eps}
\plotone{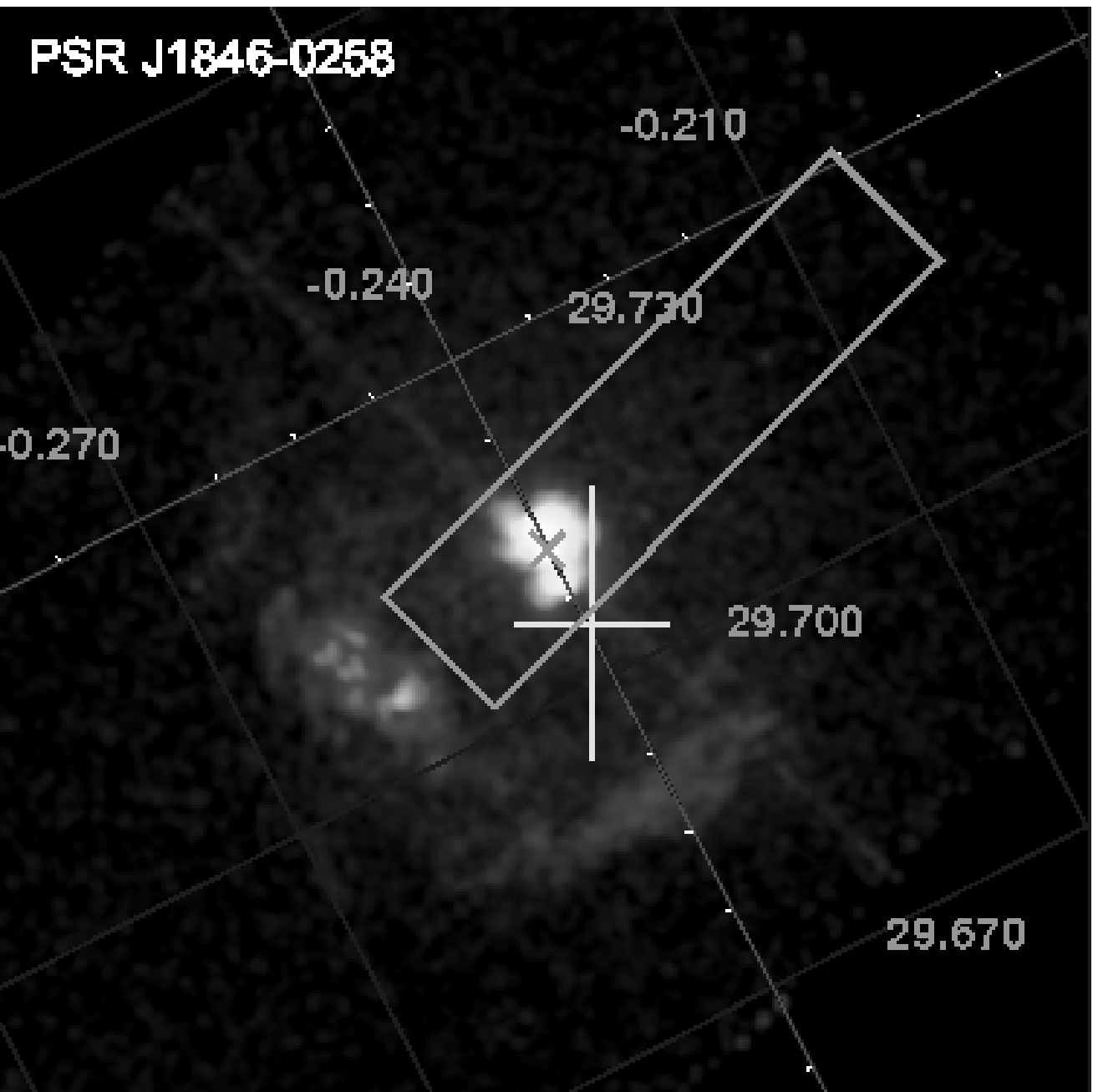}
\epsscale{.31}
%\plotone{msh1552_v1.eps}
\plotone{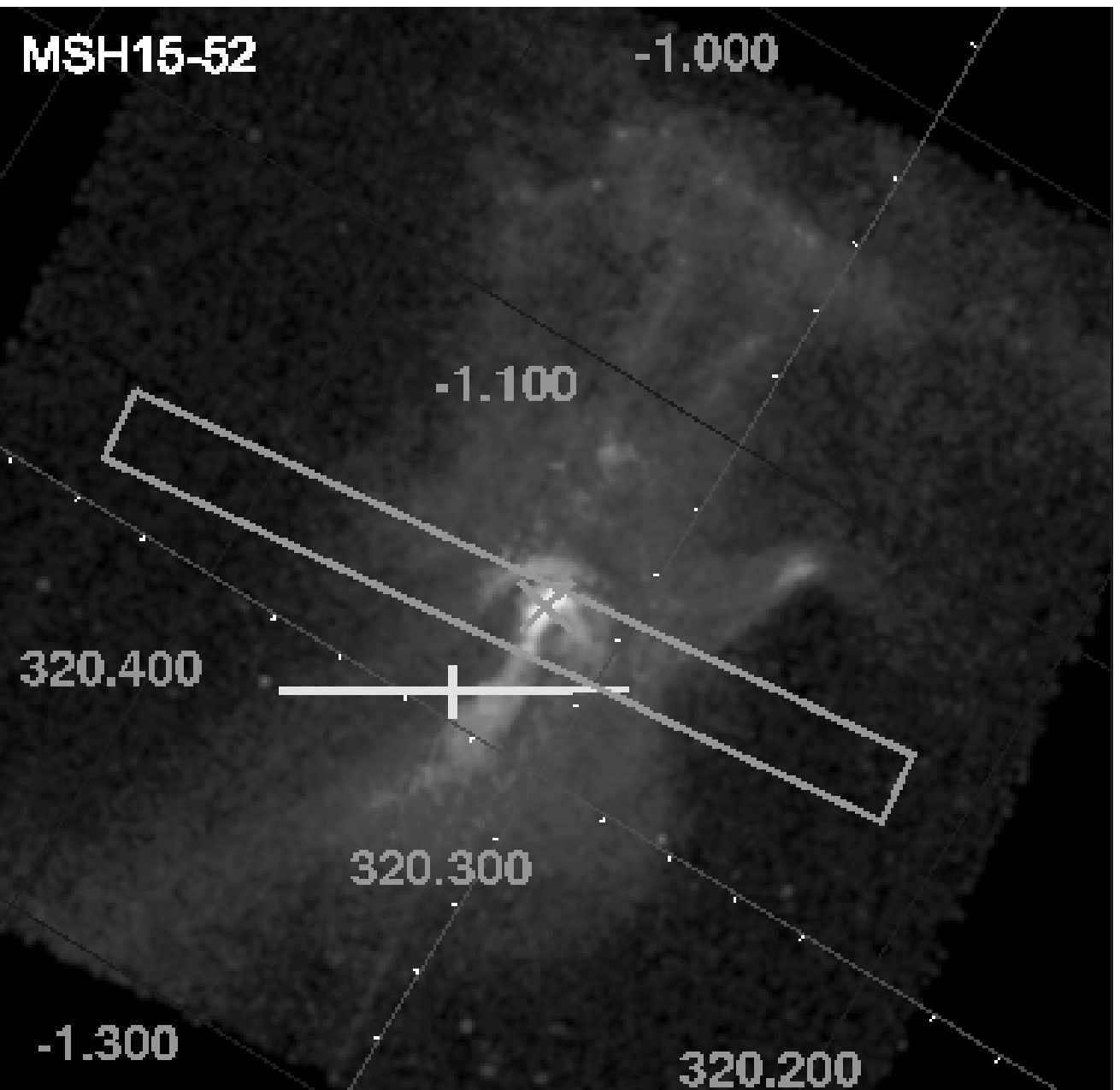}
%\plotone{G215-09_v1.eps}
\plotone{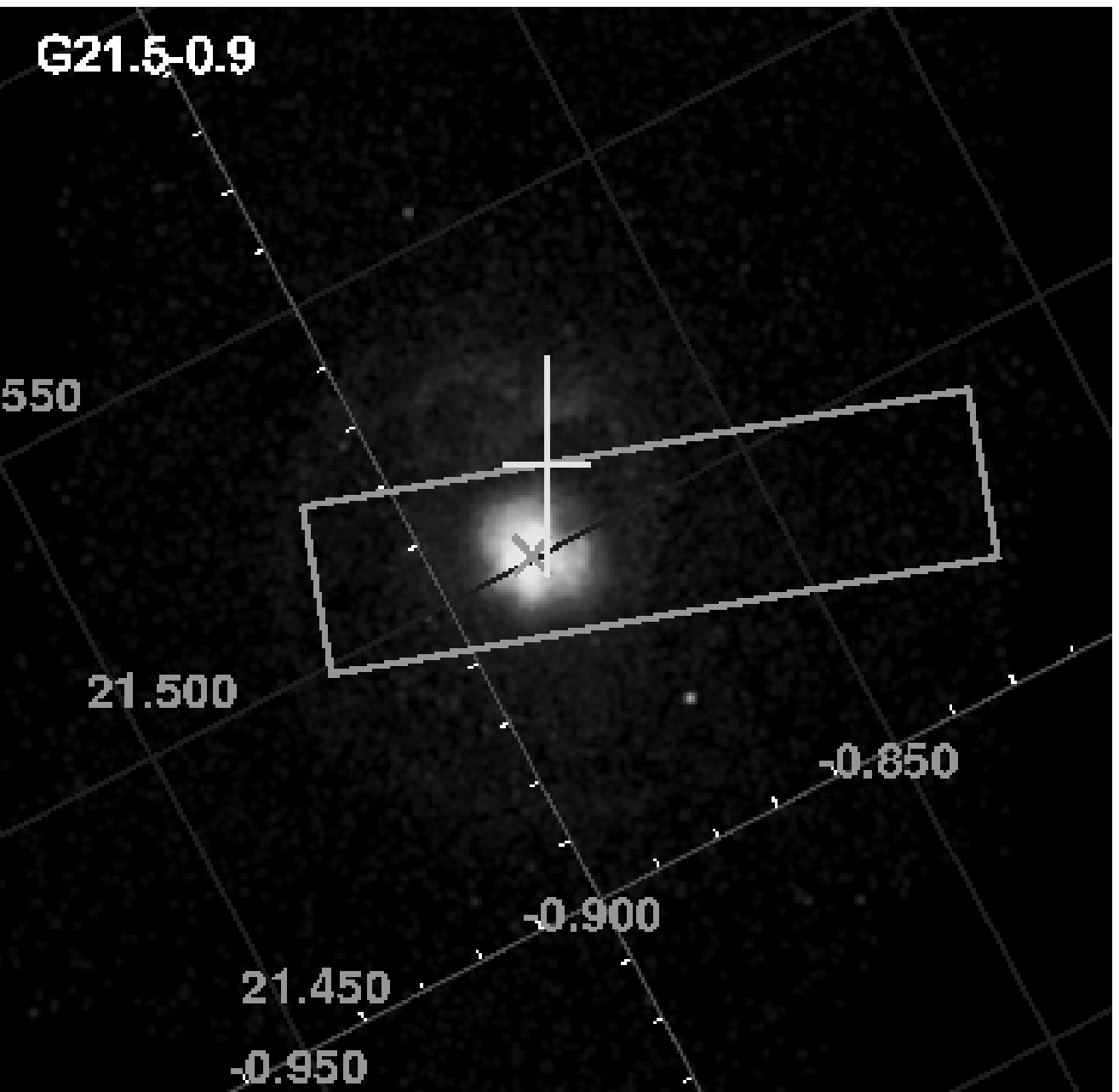}
%\epsscale{.245}
%\plotone{vela_v1.eps}
\epsscale{.31}
%\plotone{1420_v1.eps}
\plotone{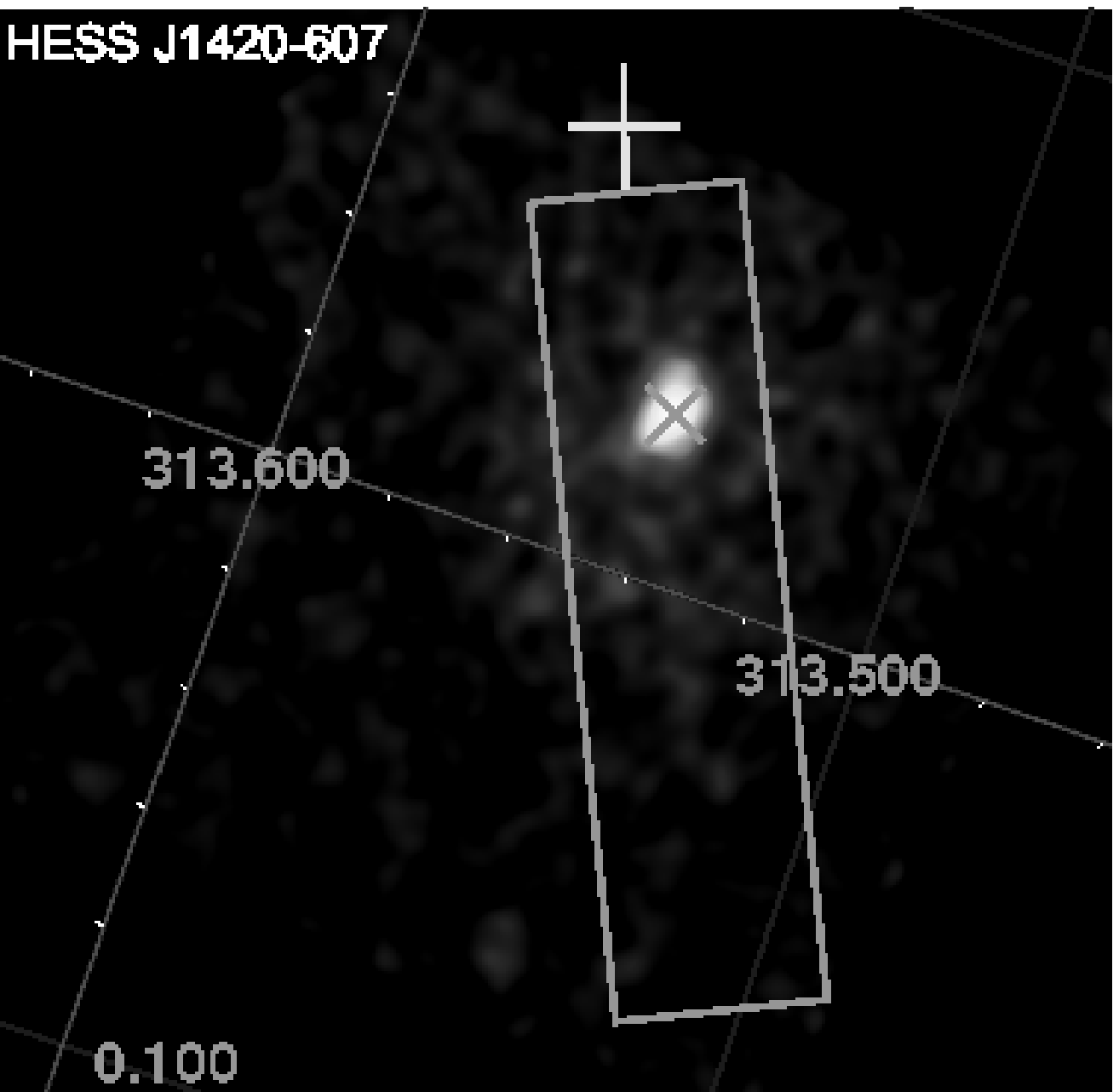}
\epsscale{.33}
%\plotone{1825_v1.eps}
\plotone{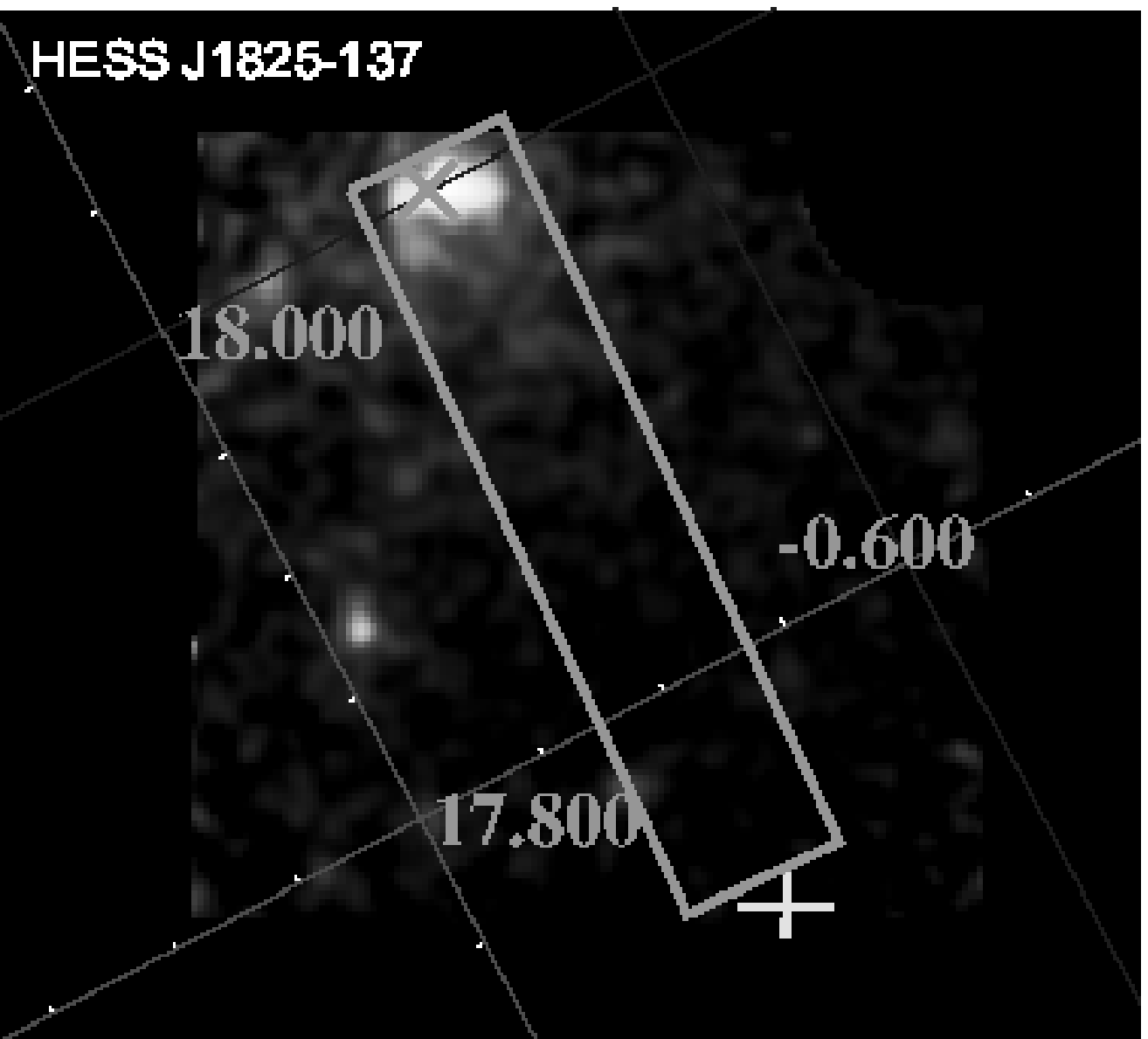}
%\plotone{1837_v1.eps}
\plotone{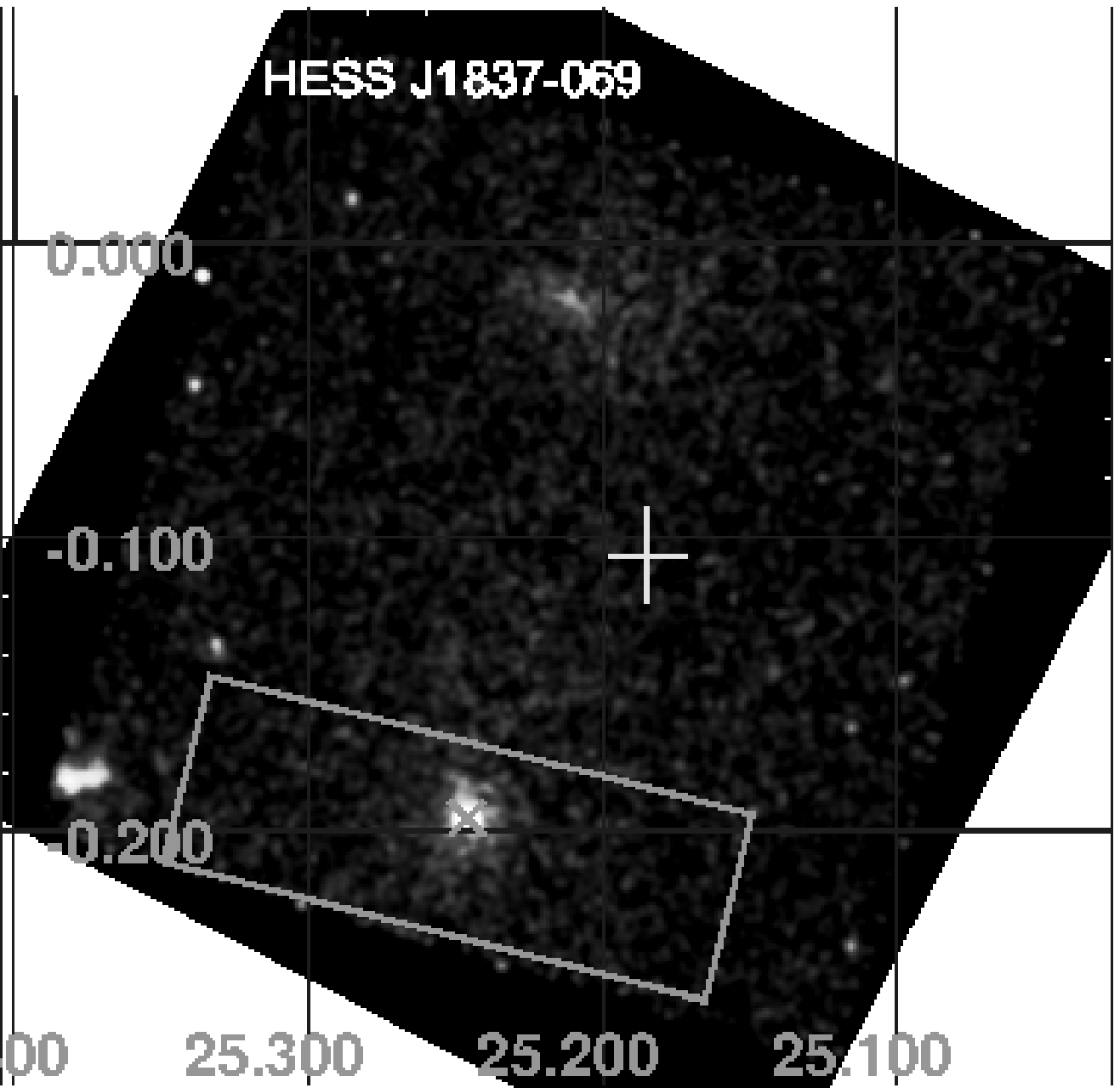}
\epsscale{.3}
%\plotone{1809_v1.eps}
\plotone{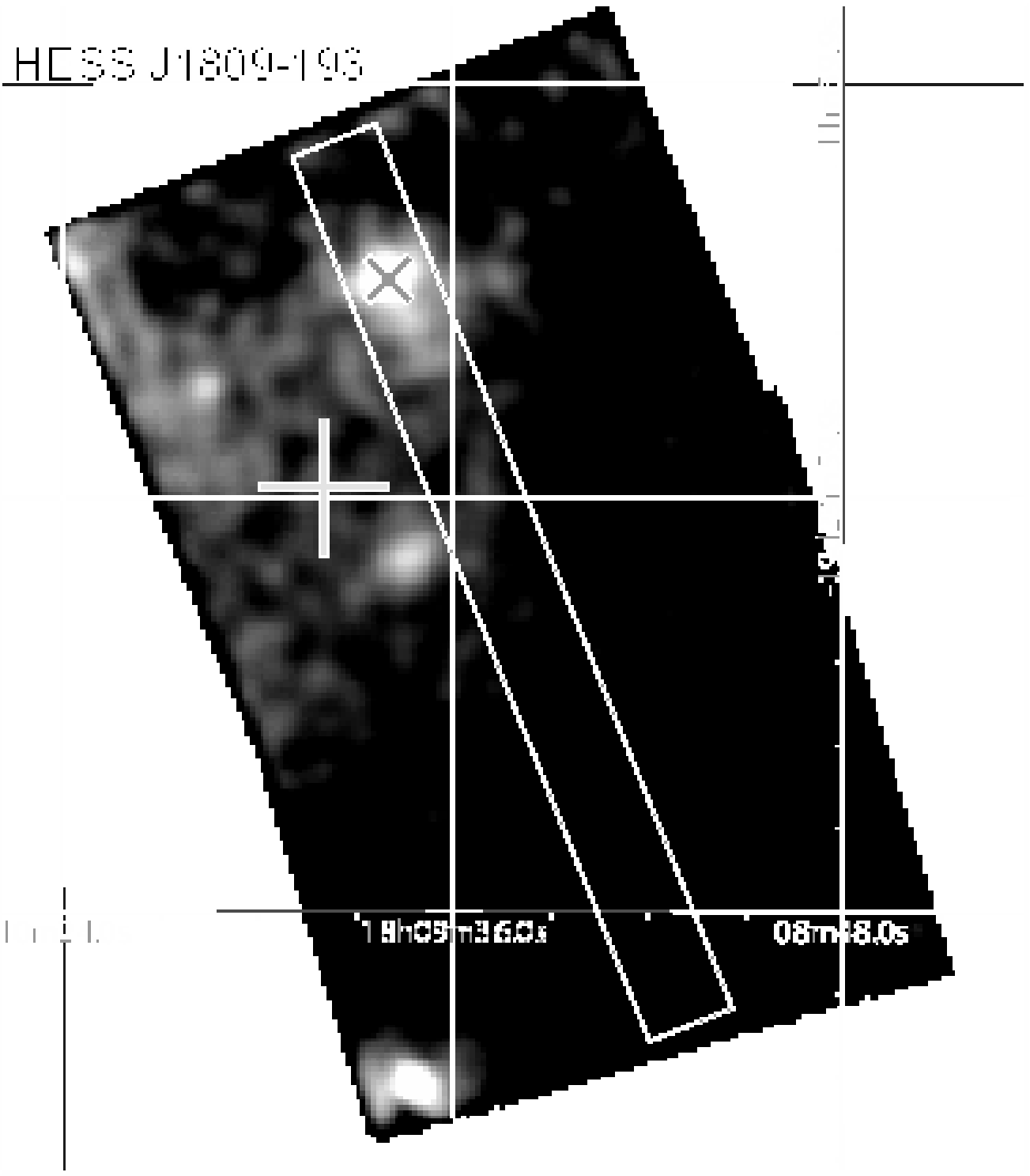}
\epsscale{.3}
%\plotone{1718_v1.eps}
\plotone{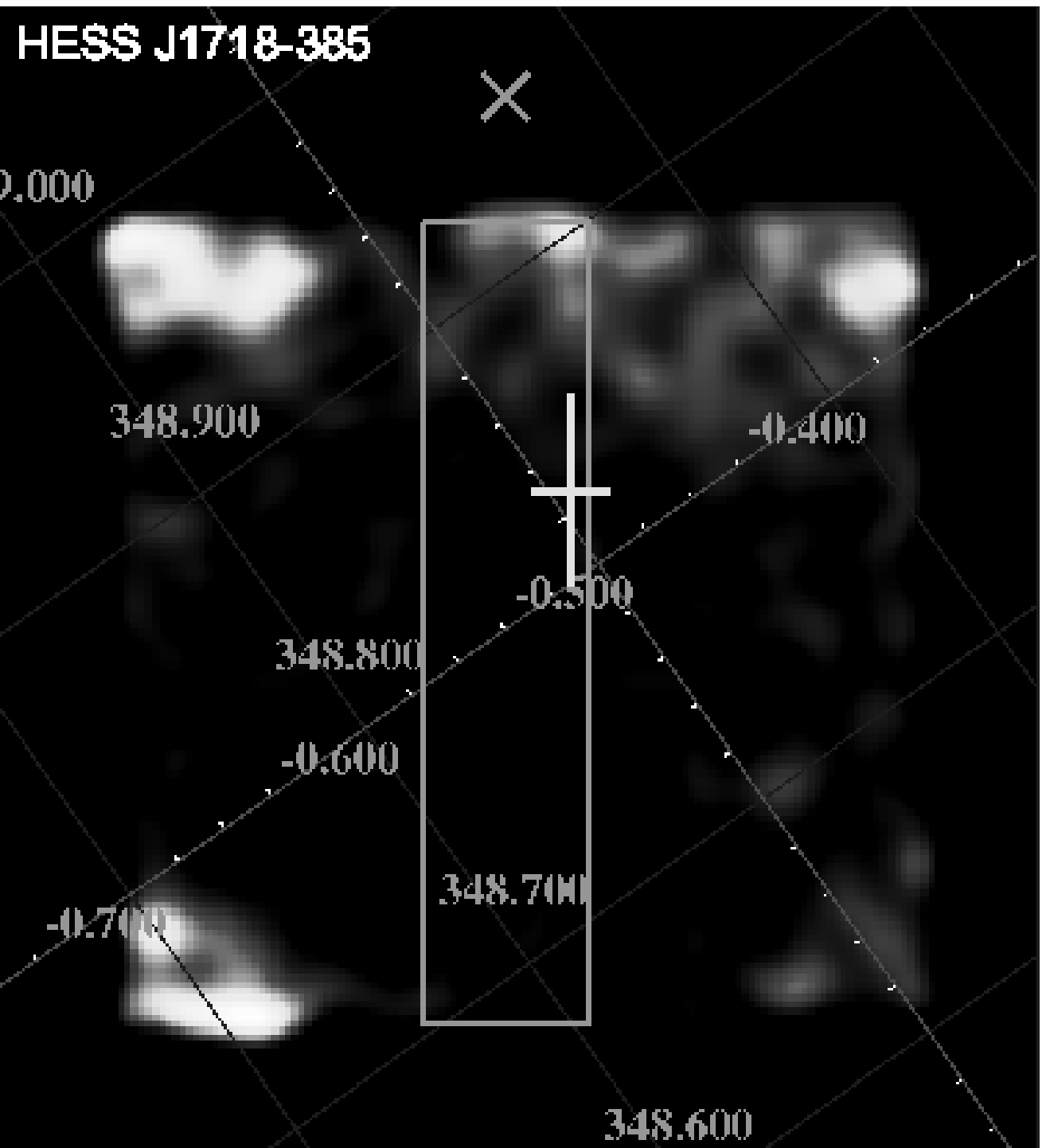}
\caption{X-ray images of the old PWNe above 2~keV
in Galactic coordinates.
The rectangles represent the region for the size estimation.
The ``+'' and ``X'' marks are the positions of VHE gamma-ray emission peaks
and the associated pulsars,
respectively.
The exposure and vignetting are corrected.}
\label{fig:image}
\end{figure}

\begin{figure}
\epsscale{.33}
%\plotone{kes75_profile.eps}
\plotone{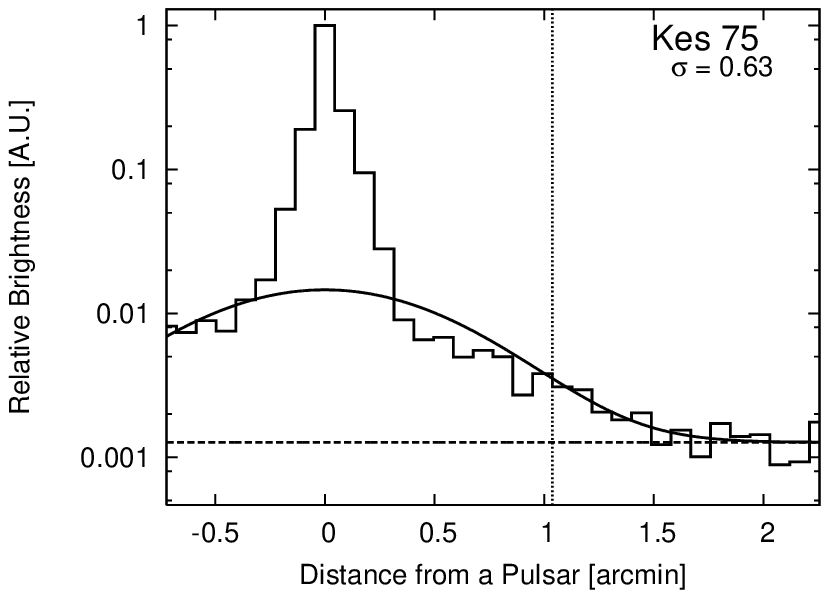}
%\plotone{msh1552_profile.eps}
\plotone{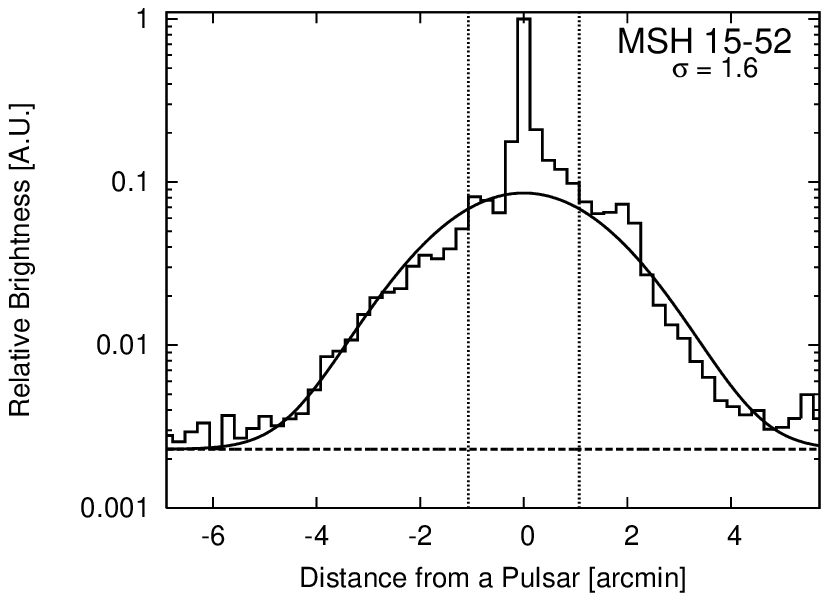}
%\plotone{G215-09_profile.eps}
\plotone{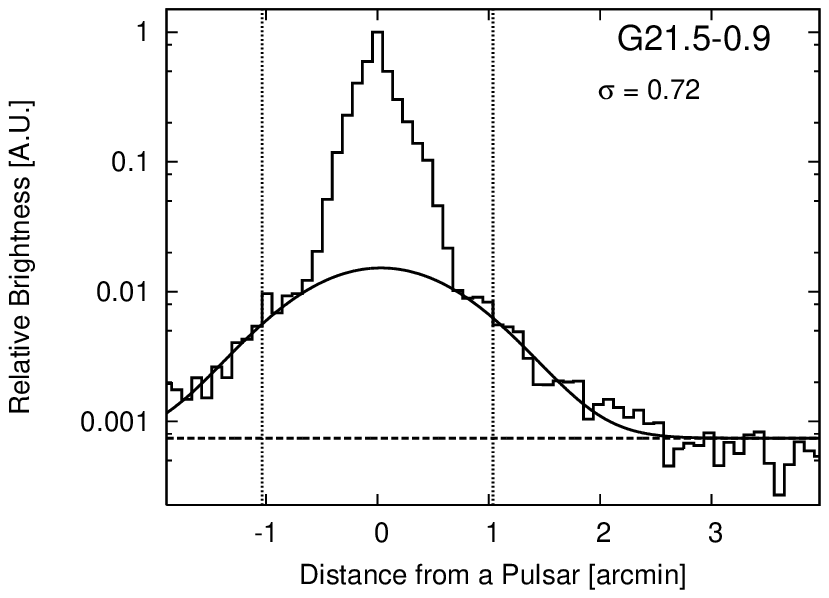}
%\plotone{vela_profile.eps}
%\plotone{1420_profile.eps}
\plotone{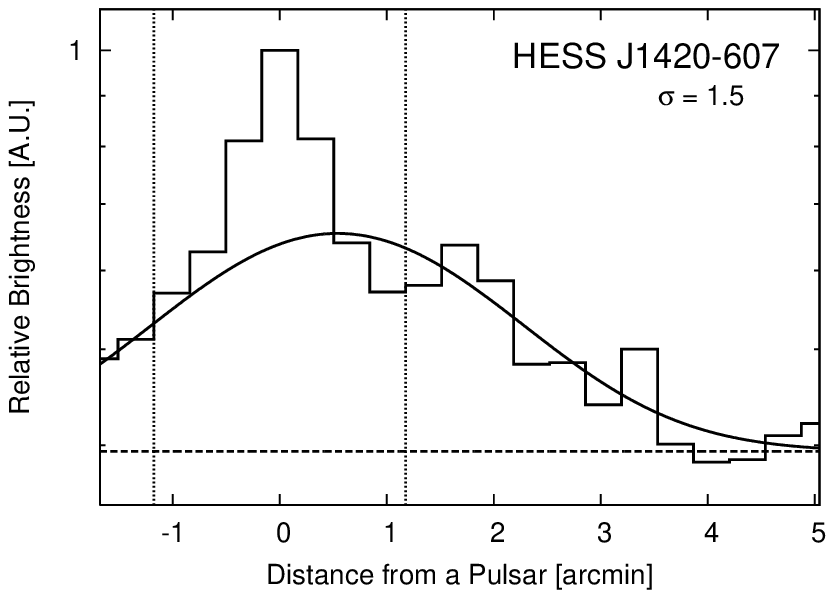}
%\plotone{1825_profile.eps}
\plotone{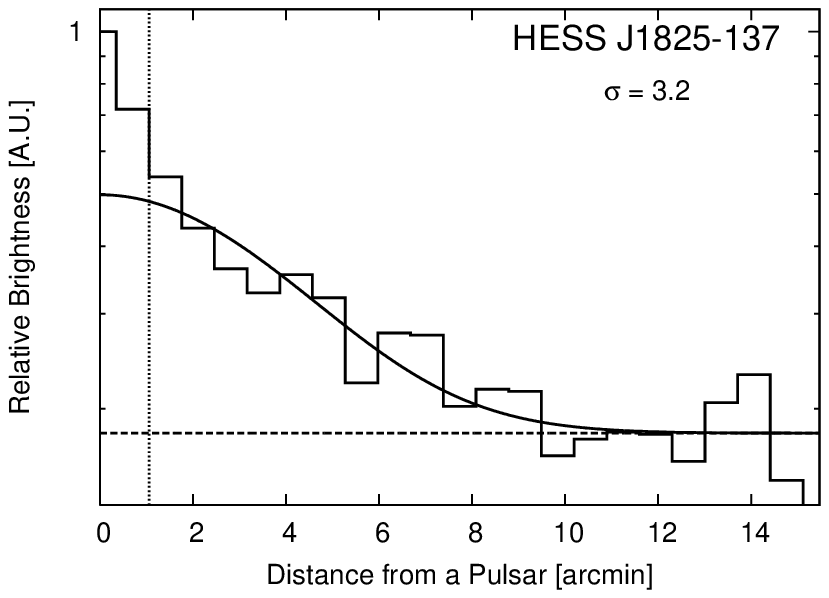}
%\plotone{1837_profile.eps}
\plotone{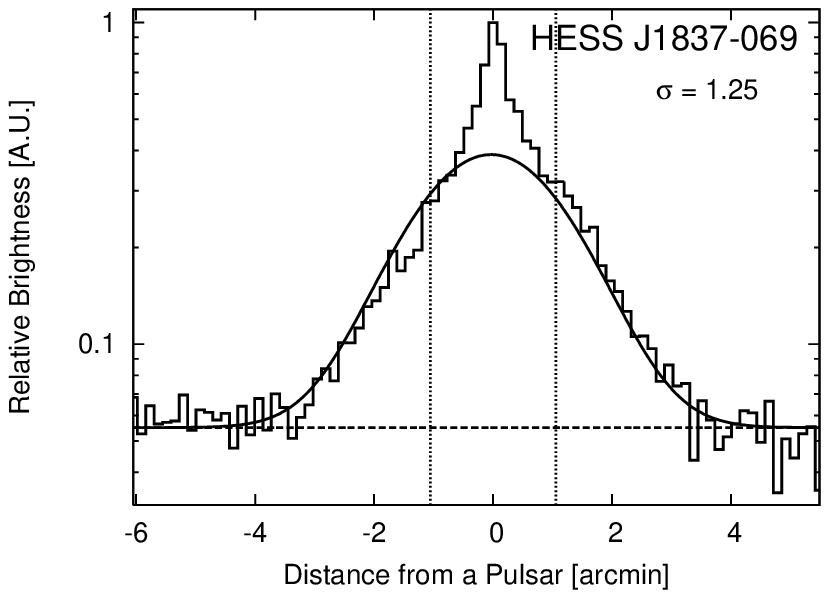}
%\plotone{1809_profile.eps}
\plotone{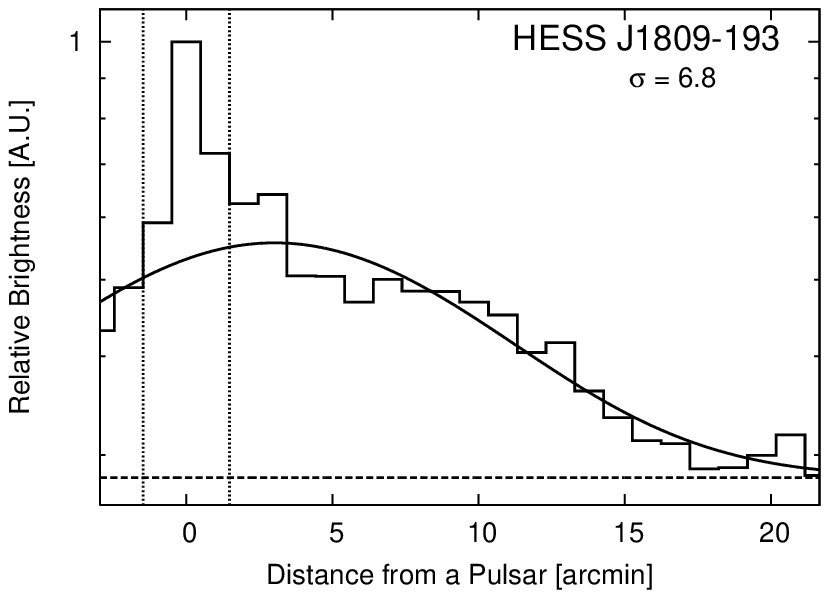}
%\plotone{1718_profile.eps}
\plotone{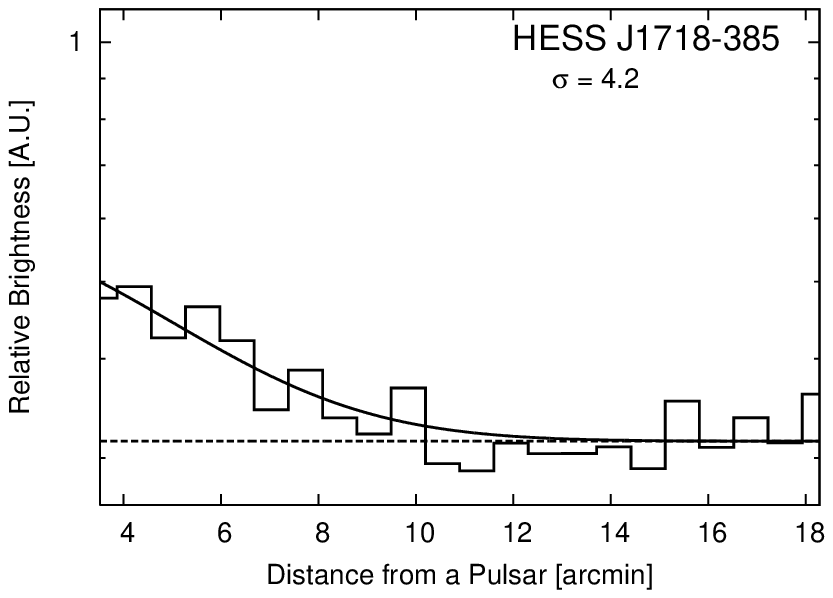}
\caption{The one-dimensional profiles of the PWNe.
The solid lines represent the best-fit gaussian model,
whereas the dashed lines are the best-fit constant background.
For the fitting, the pulsar region (surrounded by dotted lines)
was not used.}
\label{fig:profiles}
\end{figure}

Figure~\ref{fig:image} shows X-ray images above 2~keV
of our sample:
we did not use the photons below 2~keV to avoid possible contamination
of thermal X-rays from the surrounding supernova remnants.
%The non X-ray background (NXB) has been removed
%with the NXB data base \citep{tawa2008},
All images are correct for exposure and vignetting.
%For all images the exposure and vignetting effect are corrected.
%using {\tt xissim} \citep{ishisaki2007}.
As can be seen,
for the whole sample low surface brightness X-ray emission is detected
from an extended region.
%We can see largely extended emission from all the samples.
This is the first discovery of low surface brightness,
extended X-ray emission from
PWNe as old as ~100 kyr.
In order to measure the extent systematically,
we made projected profiles using
the rectangular regions shown in Fig.\ref{fig:image},
and fitted these with a 1-D Gaussian plus a constant background model
to derive the typical size of the region, using the sigma
%{\bf (or dispersion ?)
%}
of the Gaussian ($\sigma_X$).
We did not use the central pulsar region from the fitting
to avoid their influence.
As shown in Fig.\ref{fig:profiles},
a Gaussian profile provides a reasonable approximation %is effective
%except for the Vela case
for the surface brightness profiles.
%to give us rough estimation of the source sizes.}
We defined the source size as 3 times of the $\sigma_X$
in the Gaussian fit.
This is the value given in Tab.\ref{tab:result}
The best-fit parameters converted to the physical size are also shown
in Tab.\ref{tab:result}.
The errors include both the statistical errors as given by the fits,
and distance uncertainties.

%We also checked the systematic errors
%due to the direction of the profiles and the profile models
%with PSR~J1809$-$1917 and AX~J1838.0$-$0655,
%and found that the result does not show us any critical difference,
%and did not be included it.

\begin{figure}
\epsscale{.60}
%\plotone{tc_size_v2.ps}
\plotone{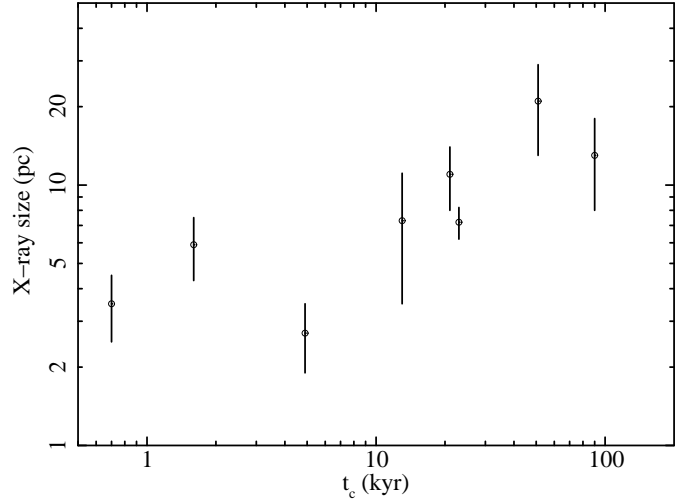}
\caption{Correlation between the characteristic age of the pulsar,$t_c$,
and the X-ray physical size of the PWNe
above 2~keV.
Errors include both statistical and distance uncertainty.
It shows evolution of the emission region size in the X-ray band.
%Although the X-ray emitting electrons lose energy typically in 2~kyr,
%size of the X-ray emission region keeps increasing well beyond a few kyr.
}
\label{fig:evolution}
\end{figure}

Figure~\ref{fig:evolution} shows the correlation
between the characteristic ages of the host pulsars
and the X-ray sizes of the PWNe. It shows that the
extent of the X-ray emission increases with 
$t_c$.
The correlation coefficient is 0.72.
The PWNe are not spherical, and as a result
the derived sizes depend on the direction of profiles.
However, the uncertainty is smaller than one order,
and the correlation cannot be removed with this effect.
If we assume that the PWN age is equal to $t_c$,
Fig.~\ref{fig:evolution} indicates that 
the X-ray emitting regions of PWNe
keep expanding up to $\sim$100~kyr.
PSR~B1706$-$44 is recently detected in VHE gamma-rays
\citep{hoppe2009},
and it has largely extended X-ray emission \citep{romani2005}.
However, no X-ray observation covered the entire emission,
and we could not measure the extension.
There are some other PWNe with known large extended X-ray nebulae,
such as PSR~J1016$-$5857 \citep{camilo2004}
and PSR~J2021+3651 \citep{vanetten2008},
from which no detection is reported in the VHE gamma-ray band.
%Although it is impossible to measure their extent
%since these observations do not cover the entire nebulae,
A rough estimation of their extent and comparing it to their
pulsar characteristic ages, are consistent with the correlation that 
we find, even though these pulsars are younger than those in our sample.
However, the sizes of these PWNe are somewhat more
uncertain,
since these PWNe are not fully covered by current X-ray observations.
Nevertheless, it suggests that
the correlation between PWNe X-ray extent and characteristic ages
is not only valid for VHE gamma-ray emitting PWNe,
but also valid for PWNe in general.

\section{Discussion}

We serendipitously found, for the first time, 
that the X-ray size of the PWNe keeps increasing 
up to $\sim$100~kyr.
In this section, we consider what makes such X-ray evolution of PWNe.

The energy loss time scale of
the electrons
emitting synchrotron X-rays with characteristic energy
$\epsilon_{syn}$ can be estimated as
\begin{equation}
2\left(\frac{B}{10\mu G}\right)^{-3/2}\left(\frac{\epsilon_{syn}}{1~{\rm 
keV}}\right)^{-1/2}~{\rm kyr\ \ ,}
\end{equation}
where $B$ is the magnetic field strength.
In the case of young PWNe such as Crab and Vela,
the magnetic field strength may be around 100~$\mu$G ---
the equipartition magnetic field strength for these PWNe 
is measured to be
60--300~$\mu$G \citep{marsden1984,hillas1998,dejager1996}.
Suppose that the X-ray emitting electrons are supplied
by the wind termination shock,
which has the radius of much less than $\sim$1~pc \citep{bamba2010}.
Then, if the old PWNe
had similar magnetic field strength of about 100~$\mu$G,
their X-ray size would be much smaller than observed
because the synchrotron cooling time scale of such electrons
would be much smaller than $\sim$ kyr.
Therefore,
in order to explain the observed correlation between 
the X-ray size and $t_c$,
the magnetic field strength inside the PWNe must decrease with time.

If the average magnetic field strength decays below 1~$\mu$G
for old PWNe,
the electrons can survive enough to emit synchrotron X-rays
with largely extended feature of $\sim20-30$~pc.
Typically, the older is
a pulsar, the smaller is its spin down luminosity.
Simultaneously the magnetic field in the PWNe may decrease,
as the magnetic field at the light cylinder decreases with time,
or because the expansion of the PWNe dillutes the magnetic field,
as the pulsar wind expands.
Alternatively, if the electron/positron energy density decreases with
time due to adiabatic expansion, and/or synchrotron cooling, and
if the magnetic field remains in rough equipartition with the
particle energy density, one also expects a decrease of the magnetic
field with time.

According to the model of \citet{kennel1984},
the electrons in PWNe are advected outwards. 
The advection speed depends on the $\sigma$ parameter,
which is defined as
the relative energy flux of magnetic field
to the kinetic energy flux.
For larger $\sigma$, the advection speed is larger,
and the electrons can escape faster.
In the case of young PWNe such as Crab and MSH~15$-$52, 
$\sigma$ is estimated to be very small, $\sim$0.003--0.03
\citep{kennel1984,mori2002,gaensler2002,yatsu2009},
although \citet{mori2004} suggested larger value ($\sigma$= 0.01--0.13).
The largely extended nebula can be achieved
if $\sigma$ becomes one order of magnitude
larger when the system becomes $\sim$100~kyr old.
%{\bf
%[Comment: $\sigma=$(magnetic energy flux)/(kinetic flux)].
%According to KC1984, wind advection speed is
%$v = [\sigma/(1+\sigma)]c$ for large radius.
%}

So far we only considered that the growth of the PWNe
is caused by advection from the expanding electron/positron plasma.
However, one should also consider the possibility of electron/positron
transport by diffusion.
For the diffusion constant $D$ one can assume
$D=\eta cE/(3eB)$, with $\eta\geq 1$ a parameter showing deviations
from Bohm diffusion ($\eta=1$). One can eliminate $E$ by
substituting the typical relation between synchrotron photon energy
and electron energy $\epsilon_{syn}=7.4 E^2B$~keV.
These can be subsituted in the
expression for the diffusion length, $l_d$ at time $t$:
$l_d = \sqrt{2Dt}$. This gives for the required magnetic field,
given $l_d$, $t$ and $\eta$:
\begin{equation}
B = \Bigl(\frac{\epsilon_{syn}}{7.4\, {\rm keV}}\Bigr)^{1/3}\Bigl(
\eta \frac{2}{3}\frac{c}{e} t
\Bigr)^{2/3}l_d^{-4/3}.
\end{equation}
This equation shows that 
in order for a PWN to reach a size of 13~pc in X-ray due to
particle diffusion, a magnetic field is required of 9~$\mu$G
(42~$\mu$G) for $\eta=1$ ($\eta=10$).
In principle the time, t, can be smaller than
the age of the pulsar,
which would correspond to a recent diffusion of the electrons,
whereas the oldest electrons have already escaped.
The estimated magnetic field strength are therefore upper-limits.
These values are not unreasonable,
but a more thorough study is required to definitely decide
whether the large X-ray extents of the PWNe in our sample are due
to diffusion or advection.
%
%It may also be important to consider the  evolution of
%the magnetic turbulence.
%The magnetic field in young PWNe may be so turbulent
%that the diffusion coefficient of particles is
%almost in the Bohm limit
%\citep[e.g.,][]{shibata2003}.
%
%The degree of magnetic field fluctuation in the PWN
%may be time-dependent.
%If this is the case,
%the diffusion coefficient of the accelerated particles
%and the mean free path of them changes with time, that is,
%accelerated particles can quickly extend to several tens of pc.
%The edge of X-ray emission could be the place
%where the diffusion become too large to keep electrons.
%When we assume that
%PWNe expand only with diffusion in 100~$\mu$G,
%the magnetic field fluctuation $\xi \equiv (/(\delta B))^2$
%becomes $\sim 4\times 10^{-2}$, which is much smaller than
%those then PWNe are young.
%The fluctuation may decay when PWNe become older,
%which is similar to the magnetic field in SNRs \citep{bamba2005}.
%More detailed models should be included as a future work.
%
%
%In reality,  
%these possibilities described above
%may be at work simultaneously.
%Typical energy of electrons emitting synchrotron X-rays ($\sim$2~keV)
%is $\sim$80~TeV with the assumption of $B = 10\mu$G.
Our result suggests that
the accelerated electrons and positrons on the termination shocks
can escape
from the host PWN systems keeping with enough energy
to emit synchrotron X-rays.
Together with the fact that the majority of 
VHE gamma-ray unidentified sources
are newly discovered old PWNe,
they could be the main contributor of
cosmic ray electrons and positrons.

Finally, we like to point out that %It is also surprising
the large extents of the PWNe of ages up
%that the PWNe keeps expanding up 
to $\sim$100~kyrs
is also surprising, because 
PWNe in a SNRs should be
crushed by the reverse shock \citep[e.g.][]{vanderswaluw2003},
whereas our result shows no evidence for 
%Our result suggests that
an influence by the reverse shock.

\acknowledgments

We acknowledge all the Suzaku team members for their gracious supports.
The authors also thank D. Khangulyan
for his comments.
This work was supported in part by
Grant-in-Aid for Scientific Research
of the Japanese Ministry of Education, Culture, Sports, Science
and Technology, No.~22684012 (A.~B.),
and No. 21740184 (R.~Y.).

\begin{deluxetable}{p{7pc}ccccccc}
\tabletypesize{\scriptsize}
\tablecaption{Observational journals and physical parameters 
for PWNe with VHE emission.
\label{tab:result}}
\tablewidth{0pt}
\tablehead{
\colhead{H.E.S.S. name} & \colhead{PWN/pulsar} & \colhead{Satellite} & 
\colhead{$t_c$\tablenotemark{a}} & \colhead{Distance} & \colhead{$\sigma_X$} & \colhead{Size} & 
\colhead{References} \\
& & & (kyr) & (kpc) & (arcmin) & (pc)
}
\startdata
HESS~J1846$-$029\dotfill & PSR~J1846$-$0258 & {\it Chandra} & 0.7 & $6.3\pm1.2$ & 
$0.63\pm0.05$ & $3.5\pm1.0$ & (1) (2) (3) \\
HESS~J1514$-$591\dotfill & MSH~15$-$52 & {\it Chandra} & 1.6 & $4.2\pm0.8$ & 
$1.6\pm0.1$ & $5.9\pm1.6$ & (4) (5) (6) \\
HESS~J1833$-$105\dotfill & G21.5$-$0.9 & {\it Chandra} & 4.9 & $4.3\pm0.9$ & 
$0.72\pm0.04$ & $2.7\pm0.8$ & (4) (2) (7) \\
%HESS~J1616$-$508\dotfill & PSR~J1617$-$5055 & {\it XMM-Newton} & 8.1 & $4.8\pm0.9$ & & 7.5 & \\
%HESS~J0835$-$455\dotfill & Vela X & {\it ASCA} & 11 & $0.29\pm0.02$ & $23.5\pm2.6$ 
%& $5.9\pm0.7$ & (8) (9) (10) \\
HESS~J1420$-$607\dotfill & PSR~J1420$-$6049 & {\it Chandra} & 13 & $5.6\pm1.1$ & 
$1.5\pm0.4$ & $7.3\pm3.8$ & (8) (4) (9) \\
HESS~J1825$-$137\dotfill & PSR~B1823$-$13 & {\it Suzaku} & 21 & $3.9\pm0.8$ & 
$3.2\pm0.2$ & $10.9\pm3.0$ & (4) (8) (10) \\
HESS~J1837$-$069\dotfill & AX~J1838.0$-$0655 & {\it Chandra} & 23 & $6.6\pm0.9$ & 
1.25$\pm$0.05 & 7.2$\pm$1.0 & (11) (12) (13)\\
HESS~J1809$-$193\dotfill & PSR~J1809$-$1917 & {\it Suzaku} & 51 & $3.5\pm0.7$ & 
$6.8\pm1.0$ & $21\pm8$ & (4) (14) (15) (16) \\
HESS~J1718$-$385\tablenotemark{b}\dotfill & PSR~J1718$-$3825 & {\it Suzaku} & 90 & 
$3.6\pm0.7$ & $4.2\pm0.5$ & $13\pm5$ & (4) (14) 
\enddata
\tablecomments{(1) \citet{leahy2008}; (2) \citet{djannati-atai2008}; 
(3) \citet{helfand2003}; 
(4) \citet{cordes2002}; (5) \citet{aharonian2005b}; (6) \citet{gaensler2002}; 
(7) \citet{slane2000}; 
%(8) \citet{dodson2003}; (9) \citet{aharonian2006b}; (10) Mori, K. et al., in prep;
(8) \citet{aharonian2006}; (9) \citet{roberts1999}; 
(10) \citet{uchiyama2009};
(11) \citet{davies2008}; (12) \citet{gotthelf2008}; (13) \citet{anada2009}; 
(14) \citet{aharonian2007}; (15) \citet{anada2010}; (16) \citet{bamba2003}; 
}
\tablenotetext{a}{The characteristic age of the pulsars.}
\tablenotetext{b}{We assumed that the center of the nebular is on the 
PSR~J1718$-$3825.}
\end{deluxetable}

\end{document}